\documentclass[12pt]{article}
\usepackage{epsfig,cite}
\usepackage{amssymb,amsmath}
\usepackage{times}

\begin{document}
\setcounter{page}{1}

\newpage
\setcounter{figure}{0}
\setcounter{equation}{0}
\setcounter{footnote}{0}
\setcounter{table}{0}
\setcounter{section}{0}


\def\theequation{\thesection.\arabic{equation}}
\def\be{\begin{equation}}
\def\ee{\end{equation}}
\def\ba{\begin{eqnarray}}
\def\ea{\end{eqnarray}}
\def\lb{\label}
\def\nn{\nonumber}

\def\a{\alpha}
\def\b{\beta}
\def\g{\gamma}
\def\d{\delta}
\def\i{\eta}
\def\e{\varepsilon}
\def\l{\lambda}
\def\r{\rho}
\def\s{\sigma}
\def\t{\tau}
\def\o{\omega}
\def\v{\varphi}
\def\x{\xi}
\def\c{\chi}

\def\D{\Delta}
\def\G{\Gamma}
\def\O{\Omega}
\def\L{\Lambda}

\def\N{\mathbb N}
\def\Z{\mathbb Z}
\def\Q{\mathbb Q}
\def\R{\mathbb R}
\def\C{\mathbb C}
\def\F{\mathbb F}
\def\H{\mathbb H}

\def\E{{\cal E}}

\def\id{\mbox{\rm 1\hspace{-4.5pt}I}}

\def\p{\hat p}

\def\bo{\mathfrak b}

\def\tp{\tilde p}

\def\Rp{\hat{R}(p)}

\def\bU{\overline{U}_q}
\def\tU{{\tilde{U}}_q}
\def\bD{\overline {\cal D}}

\def\uz{\underline z}

\def\eod{\phantom{a}\hfill \rule{2.5mm}{2.5mm}}


\title{Extended $\widehat{su}(2)_k$
and restricted $U_q sl (2)$}

\author{Ludmil Hadjiivanov$^{a,c,d~1},\ $ Paolo Furlan$^{b,c~2}\ $ }

\date{}

\maketitle
\vspace{0.3cm}

\hfill{\em To Yvonne\qquad}

\vspace{1.3cm}

\scriptsize{
$^a$ International School for Advanced Studies
(SISSA/ISAS), via Beirut 2-4, I-34014 Trieste, Italy

$^b$ Dipartimento di Fisica Teorica dell'
Universit\`a degli Studi di Trieste, Strada Costiera 11, I-34014
Trieste, Italy

$^c$ Istituto Nazionale di Fisica Nucleare (INFN), Sezione di
Trieste, Trieste, Italy

$^d$ Theoretical and Mathematical Physics Division, Institute for
Nuclear Research and Nuclear Energy,
Bulgarian Academy of Sciences, Tsarigradsko Chaussee 72, BG-1784 Sofia, Bulgaria }

\vspace{0.3cm}

\begin{abstract}

{Global gauge symmetry becomes more intricate in low dimensional QFT. We survey the
mathematical concepts leading to the relevant analogues of the ($D=4$)
Doplicher-Haag-Roberts theory of superselection sectors and internal symmetry.
We also review a recently uncovered duality between braid and quantum group representations
in an extension of the chiral $\widehat{su}(2)_k$ WZNW model for nonnegative integer level $k$.}

\end{abstract}

\vfill

\footnoterule \footnotesize{
$^1$ e-mail address: lhadji@inrne.bas.bg

$^2$ e-mail address: furlan@trieste.infn.it}

\newpage


\section{Tannaka-Krein, Doplicher-Roberts and Kazhdan-Lusztig}


The existence of charges, i.e., quantities that are conserved independently of
the concrete dynamics and hence, are represented by operators commuting not only with
the Hamiltonian but with all the observables is reflected, quantum field theoretically,
in the notion of {\it superselection sectors}, the charge eigenspaces. Localizable charges generate {\it internal (gauge) symmetry} which is, in turn, intimately related to {\it statistics}. In space-time dimension $D\ge 3+1$ this is accounted for, in the framework of the algebraic (local, relativistic) QFT \cite{Haag}, by the Doplicher-Roberts (DR) theorem \cite{DR}  implying that, typically,  the full Hilbert space ${\cal H}$ of the theory decomposes in terms of the superselection sectors ${\cal H}_p$, inequivalent representations of the algebra of observables generated from the vacuum by (unobservable) charged fields, as
\be
{\cal H} = \bigoplus_p\, {\cal H}_p \otimes {\cal V}_p\ ,\qquad d_p:= \dim {\cal V}_p < \infty\ ,
\lb{HQFQ}
\ee
where ${\cal V}_p$ are (finite dimensional ) representations of a {\it compact} gauge group $G$ whose action
leaves the observables invariant. The exchange properties of the charged field generating a sector are characterized by a {\it statistics parameter} $\l_p = \pm \frac{1}{d_p}$. If the {\it statistics dimension} $d_p$ is equal to $1$, the sign factor reflects the usual Bose-Fermi alternative. In general, higher dimensional representations of the permutation group are admitted which correspond to (Bose or Fermi) parastatistics; in this case, the integer $d_p$ is its order. As it follows from (\ref{HQFQ}) and the Clebsch-Gordan decomposition of $G$, the statistics dimensions obey {\it fusion rules} of the type
\be
d_{p_1}  d_{p_2} = \sum_p \, N_{p_1 p_2}^p\, d_p\ ,\qquad N_{p_1 p_2}^p \in {\Z}_+\ .
\lb{fr}
\ee

Formula (\ref{HQFQ}) is reminiscent of the classical Schur-Weyl duality, where
the $k$-th tensor power of the defining representation of $GL(N,{\C})$ (on which the
permutation group ${\mathfrak S}_k$ acts by exchanging the order of factors) decomposes in a sum,
over the set of partitions $Part (k,N)$ of $k$ in not more than $N$ parts,
of tensor products of irreps of $GL(N,{\C})$ and ${\mathfrak S}_k$, the corresponding Young diagram of $k$ boxes in $r\ (\le N)$ rows labeling, in the first case, the highest weight:
$({\C}^N )^{\otimes k} \simeq \oplus_{Y\in Part (k,N)}\, T_Y \otimes V_Y\ .$
Thus, the endomorphisms coming from the one group centralize (the group algebra of) the other.
Similar dualities exist for the symplectic and the orthogonal groups, with the group algebra of the permutation group replaced by the corresponding Brauer algebra, and also for some $q$-deformations (with $q$ generic) where this role is played by the Hecke (in the type A case), or Birman-Murakami-Wenzel (BMW) algebras, respectively, see e.g. \cite{Doty}.

The DR theorem establishes the equivalence of two different representation categories --
the one of charge endomorphisms of the algebra of local observables, and that of compact groups.
(In particular, the two sets of representations are identical and hence, can be parametrized by
the same labels.) The DR equivalence is considered as a non-commutative $C^*$-algebraic generalization of the {\it Tannaka-Krein duality} between compact groups and the category of their finite dimensional representations which, in turn, generalizes the Pontryagin duality between abelian compact groups and their characters to the non-abelian case.

For $D = 1+1$ (and, in the case of non-localizable charges, also for $D = 2+1$), the more involved causal
space-time structure leads to path depending exchange factors and, correspondingly, to {\it braid group} representations.\footnote{Following F. Wilczek, fields corresponding to $1$-dimensional braid representations are called "{\it anyons}". Those obeying non-abelian braid statistics are sometimes
referred to as "{\it plektons}", from the greek word for braid.}
This leads, in turn, to other drastic changes: the phase of the statistics parameter $\l_p$ may be non-trivial, and the statistics dimension $d_p\ge 1$ may take non-integer values \cite{FRS}. The latter fact alone rules out the possibility of having a gauge symmetry of group type. This role is now taken by a "quantum group" (QG) \cite{TK}, an algebra of Hopf type with some additional structures.
For a recent review of the achievements in classifying representations of local conformal nets of von Neumann factors, see e.g. \cite{YK}.

The best studied class of two dimensional QFT is that of {\it rational} conformal field theories (RCFT) for which the category of representations of the vertex operator algebra (VOA; the analog of the algebra of observables) is {\it semisimple}, has {\it finitely many} simple objects (equivalence classes of irreducibles), and obeys certain non-degeneracy requirements i.e., is a {\it modular} tensor category (see the excellent book of B. Bakalov and A. Kirillov, Jr. \cite{BK}). It has been proven quite recently, see \cite{Hayashi, O1}, that a finite semisimple tensor category is equivalent to the representation category of a {\it weak Hopf algebra} \cite{BS, BNS} or, alternatively, of a  related Ocneanu's {\it double triangle algebra}, see e.g. \cite{PZ} and references therein.

The extension of the VOA-QG correspondence to finite but not necessarily semisimple categories is
under intensive study, both by mathematicians and mathematical physicists, see \cite{EO} and \cite{F1}. Non-semisimple fusion algebras appear e.g. in {\it logarithmic} conformal theories (LCFT), in particular in some logarithmic extensions of minimal models that have been studied previously by H.G. Kausch, M.R. Gaberdiel, M. Flohr, etc. A general classification seems out of reach, so it is worth studying reasonable subclasses.

D. Kazhdan and G. Lusztig (KL) have established, in the series of papers \cite{KL}, the equivalence of certain tensor category of representations of the affine algebra $\hat{\mathfrak g}_{h-{g^\vee}}$ at {\it height} $h\notin{\Q}_{\ge 0}$ with that of the finite dimensional modules of the quantum universal enveloping algebra (QUEA) $U_q({\mathfrak{g}})$ at $q= e^{i\frac{\pi}{mh}}$ (where $m=1$ for simply laced ${\mathfrak g}$).\footnote[2]{Here $g^\vee$ is the dual Coxeter number of the simple Lie algebra ${\mathfrak g}$. Note that $h$ is {\it allowed} to take negative rational values, $h\in{\Q}_{< 0}$, when $q$ is a root of unity and both categories are non-semisimple.} Such a direct relation between the {\it algebraic objects} $\hat{\mathfrak g}_k$ and $U_q({\mathfrak{g}})$ (through their representation categories) is missing in the semisimple case of {\it integrable} ${\hat{\mathfrak g}}_k$ modules, for $k\in {\Z}_+$, where
the equivalence is obtained only after taking the quotient with respect to an ideal of indecomposable modules of $U_q({\mathfrak g})$.\footnote[3]{Cf. \cite{PS, FK}. The precise construction of the $U_q({\mathfrak{g}})$ category uses the notion of {\it tilting modules} \cite{AP}. One of the proofs (see \cite{BK}) of its equivalence with the category of integrable $\hat{\mathfrak g}_{h-{g^\vee}}$ modules has been given by M. Finkelberg \cite{Fin} who combined results of KL with certain duality $h\leftrightarrow -h$.}

It has been shown recently, in another series of papers by B. Feigin et al. \cite{FGST}, that a KL correspondence exists between the VOA of a $(p, p')$ LCFT model and certain (finite dimensional, factorizable, ribbon) Hopf algebra, in the sense that the corresponding representation categories,
fusion and modular properties are equivalent.
In particular, a KL correspondence has been established, in the first paper of \cite{FGST}, between the $(1,h)$ LCFT model ($h\ge 2$) and the $2h^3$-dimensional {\it restricted} QUEA $\bU \equiv \bU s\ell(2)$ for $q=e^{\pm \frac{i\pi}{h}}\, $.  The latter is generated by $E, F, q^{\pm H}$, satisfying
the relations
\ba
&&q^{\pm H} q^{\mp H} = \id\ ,\qquad q^H E  = q^2 E q^H\ ,\qquad q^H F  = q^{-2} F q^H\ ,\qquad\nn\\ &&[E,F]=[H]:=\frac{q^H-q^{-H}}{q-q^{-1}}\ ,\qquad
E^h = 0 = F^h\ ,\quad (q^H)^{2h} = \id\ .\qquad\qquad
\lb{bU}
\ea
The Hopf structure (coproduct $\Delta$, counit $\e$ and antipode $S$) on it is given by
\ba
&&\Delta (E) = E\otimes q^H + \id\otimes E\ ,\quad \Delta (F) = F\otimes\id + q^{-H}\otimes F\ , \nn\\
&&\Delta (q^{\pm H}) = q^{\pm H} \otimes q^{\pm H}\ ,\qquad \e (E) = 0 = \e (F)\ ,\quad \e (q^{\pm H}) = 1\ ,\nn\\
&&S (E) = - E q^{-H}\ ,\quad S (F) = - q^H F\ , \quad S (q^{\pm H}) = q^{\mp H}\ .
\lb{HopfbU}
\ea

We shall review in what follows some results of \cite{FHT7} signaling a similar relation
between a logarithmic-type extension of the ${\widehat{su}}(2)_{h-2}$ chiral WZNW model for integer $h\ge 2$
and Lusztig's extension $\tU$ of $\bU$ at $q=e^{\pm i\frac{\pi}{h}}$.

\section{Braid representations on the regular solutions of the $\mathbf{{\widehat{su}}(2)_{h-2}}$ KZ equations}


For a semisimple Lie algebra ${\mathfrak g}$, the Knizhnik-Zamolodchikov system of linear partial differential equations reads
\be
\lb{KZ4}
\left( h\frac{\partial}{\partial z_a} - \sum_{{b=1}\atop{b\ne a}}^N \frac{C_{ab}}{z_{ab}}\right) w (z_1,\dots ,z_N) = 0\ ,
\quad a =1,\dots ,N\ ,\quad z_{ab}=z_a-z_b\ ,
\ee
where $C_{ab}$ is the polarized quadratic Casimir operator acting on the tensor product of ${\mathfrak g}$-modules attached to $a$ and $b$ (so that, in particular, $C_{ab} =C_{ba}$ and $[C_{ab} , C_{ac} + C_{bc} ] = 0$ for distinct $a,b,c$).
In the case when $w$ is a chiral block of $N=4$ WZNW primary fields, M\"obius (projective) invariance dictates
that $w (\uz)$ is a function of the harmonic ratio $\eta = \frac{z_{12}z_{34}}{z_{13}z_{24}}$ times a scalar
prefactor, depending on $z_{ab}$ only\footnote[4]{There is an $\eta$-dependent multiplicative freedom in choosing the prefactor.}. On the other hand ${\mathfrak g}$-invariance, implying $( \sum_{b\ne a} C_{ab} + C_a)\,w (\uz) = 0$, where
$C_a$ is the Casimir eigenvalue in the $a$-th ${\mathfrak g}$-module, reduces the number of independent terms
$C_{ab} w (\uz)\,,\ 1\le a< b\le N\,,$ from $\left( N\atop 2\right)\,$ to $\frac{N(N-3)}{2}$.
Restricting our attention to the $SU(2)$ WZNW model at level $k=h-2\in {\Z}_+$,
we shall make use of the polynomial realization of the $su(2)$ modules, introducing auxilliary complex variables
$\zeta_a\,,\ a=1,\dots ,4$, so that the Casimir operators become second order differential operators in them,
cf. \cite{FZ, FGPP, STH}. Finally, for the four-point chiral block $w^{(p)} ({\underline\zeta}, \uz)$ of a single primary field of
shifted weight $p\in{\N}$ (i.e., of isospin $I$ such that $p=2I+1$,
and conformal dimension $\Delta_p= \frac{p^2-1}{4h}$), the KZ system (\ref{KZ4}) reduces to
\be
\left( h \frac{\partial}{\partial \eta} - \frac{\Omega^{(p)}_{12}}{\eta} + \frac{\Omega^{(p)}_{23}}{1-\eta} \right)\,  f^{(p)}(\xi , \eta) = 0\ ,
\lb{KZ4f}
\ee
where $\xi = \frac{\zeta_{12}\zeta_{34}}{\zeta_{13}\zeta_{24}}$ and $f^{(p)}$ is a polynomial in $\xi$
of order $p-1$ such that
\ba
&&w^{(p)} ({\underline\zeta}, \uz)
= \left( \frac{z_{13}z_{24}}{z_{12}z_{23}z_{34}z_{14}} \right)^{2\Delta_p}
(\zeta_{13}\zeta_{24})^{p-1}\,  f^{(p)}(\xi , \eta) \ ,\nn\\
&&\Omega^{(p)}_{12} = \Omega^{(p)}(\xi , \frac{\partial}{\partial\xi})\ ,\qquad \Omega^{(p)}_{23} = \Omega^{(p)} (1-\xi , - \frac{\partial}{\partial\xi})\ ,\lb{w-f}
\ea
$$\Omega^{(p)}(\xi , \frac{\partial}{\partial\xi}) = (p-1)(p-(p-1)\,\xi)-(2(p-1)-(2p-3)\,\xi)\,\xi\frac{\partial}{\partial\xi}+\xi^2(1-\xi)\frac{\partial^2}{\partial\xi^2}\ .$$

A set of $p$ linearly independent solutions $\{ f^{(p)}_\mu (\xi, \eta) \}_{\mu = 0}^{p-1}$ of Eq. (\ref{KZ4f}) has been constructed explicitly in terms of multiple integrals
in \cite{STH} for any $p=1,2,\dots$ . They span a representation ${\cal S}_p$ of the braid group with generators $b_i$ corresponding to the exchange of variables with labels
$i$ and $i+1$ in $w^{(p)}({\underline\zeta}, \uz)$ (\ref{w-f}), the homotopy class of the exchange
of points being fixed so that $z_{i\, i+1}\rightarrow e^{-i\pi} z_{i\, i+1}$. In terms of
$f(\xi, \eta)$, the braidings $b_i$ act as\footnote[5]{The actions of $b_1$ and $b_3$ coincide, so one deals, effectively, with the braid group ${\mathfrak B}_3$.}
\ba
&&b_1  f^{(p)}_\mu (\xi, \i)\,=\, (1-\xi)^{p-1}(1-\i)^{4\Delta_p}
f^{(p)}_\mu(\frac{\xi}{\xi-1},\frac{e^{-i\pi}\i}{1-\i}) = f^{(p)}_\l(\xi,\i)\,{B_1}^\l_{~\mu}\ ,\nn\\
&&b_2 f^{(p)}_\mu (\xi, \i)\, = \, \xi^{p-1}\i^{4\Delta_p}
f^{(p)}_\mu(\frac{1}{\xi},\frac{1}{\i}) = f^{(p)}_\l(\xi,\i)\,{B_2}^\l_{~\mu}\ ,
\lb{b1b2}
\ea
respectively. Here $({B_1}^\l_{~\mu})$ and $({B_2}^\l_{~\mu})\,,\ \l\,,\, \mu = 0,1,\dots , p-1$ are
(lower and upper, respectively) triangular $p\times p\,$ matrices:
\be
{B_1}^\l_{~\mu} =  (-1)^{p-1-\l} q^{\l (\mu +1)-\frac{p^2-1}{2}}
\left[{\l\atop\mu}\right] = {B_2}^{p-1-\l}_{~p-1-\mu}\ .
\lb{B1B2}
\ee
Due to the fact that this set of solutions is well defined also beyond the integrability bound $p=h-1$
(where ``unitary'' bases become singular), it has been called in \cite{STH} ``the regular basis''.
The Gaussian (or $q$-)binomial coefficients above are defined for any integer $a$ and non-negative integer $b$ as
\be
\left[{a\atop b}\right] := \prod_{t=1}^b \frac{q^{a+1-t}-q^{t-a-1}}{q^t-q^{-t}}\ ,\quad b\ge 1\ ,\qquad \left[{a\atop 0}\right] := 1\ .
\lb{qbin}
\ee
It follows \cite{L} that $\left[{a\atop b}\right]\in{\Z}[q,q^{-1}]$, and
\begin{equation}
\left[{a\atop b}\right] = 0\quad{\rm if}\quad 0\le a<b\ ,\qquad \left[{a\atop b}\right] = \frac{[a]!}{[b]![a-b]!}\quad{\rm for}\quad 0\le b\le a\ .
\label{qbin2}
\end{equation}

Primary fields of integer isospin and conformal dimension are local (also with respect to themselves) if and only if their $4$-point function is rational. Due to the special choice of the prefactor in (\ref{w-f}), the rationality of $w^{(p)}({\underline\zeta}, \uz)$ implies that $f^{(p)}(\xi, \eta)$ is a polynomial (of order not exceeding $4 \Delta_p$) also in $\eta$. The list of polynomial solutions of (\ref{KZ4f}) reproduces, for $I\le \frac{k}{2}$ (or, equivalently, $p\le h-1$), the ADE classification of the local extensions of the $\widehat{su}(2)_k$ current algebra \cite{MST}.

Braid invariant polynomial solutions $f^{(2h-1)}_{h-1}$ have been explicitly constructed later in \cite{HP} also for $p=2h-1$ (corresponding to isospin $I=k+1$), for any non-negative integer level. They do not obey the integrability condition, so the corresponding local primary field of integer conformal dimension $\Delta_{2h-1} = h-1$ should give rise to a non-unitary representation of the $\widehat{su}(2)_k$ current algebra. It has been noticed further by A. Nichols \cite{N1} that, in fact, for any $p=2(J+1)h-1\,,\, J=0,\frac{1}{2},1,\dots$ the $(2J+1)$-dimensional subspace of ${\cal S}_p$ spanned by $\{ f^{(p)}_{mh-1} \}_{m=0}^{2J}$ forms an irreducible representation of the braid group under the action defined in (\ref{b1b2}) (the invariant found in \cite{HP} corresponds to the singlet $J=0$).

We shall display, as an example, the regular basis for $h=2\,,\, p=2h-1=3$. The general formula in \cite{HP} for the polynomial invariant reduces in this case to
\be
f_1^{(3)}(\xi,\eta) = \eta(1-\eta)(\eta(1-2\xi)-\xi(\xi-2))\ ,
\lb{p3poly}
\ee
while the two other regular basis solutions of (\ref{KZ4f}) are logarithmic:
\ba
&&f_0^{(3)}(\xi,\eta) = -\frac{1}{\pi} ( f_1^{(3)}(\xi,\eta) \ln\eta + (1-\eta)^2 (\eta^2-\xi^2))\ ,\nn\\
&&f_2^{(3)}(\xi,\eta) = f_0^{(3)}(1-\xi,1-\eta) \ .
\lb{p3log}
\ea
One can easily check that (\ref{b1b2}) holds in this case with $q=e^{-i\frac{\pi}{h}}=-i$ and
matrices $B_1 ,B_2$ as in (\ref{B1B2}),
\begin{equation}
B_1=\left(\begin{matrix}1&0&0\\i&1&0\\-1&0&-1\end{matrix}\right)\ ,\quad
B_2=\left( \begin{matrix}-1&0&-1\\ 0&1&i\\ 0&0&1\end{matrix} \right)\ .
\label{BBp3}
\end{equation}

The structure of the spaces ${\cal S}_p$ as braid group modules has been studied in full generality
in \cite{FHT7} and is the following. Let $1\le r\le h-1$ and $N\ge 1$ be both integer; then,
all ${\cal S}_r$ as well as ${\cal S}_{Nh}$ are irreducible, while each ${\cal S}_{Nh+r}$
contains an $N(h-r)$-dimensional invariant irreducible submodule $S_{N , h-r}$ such that
the corresponding $(N+1)r$-dimensional quotient ${\tilde S}_{N+1, r}$ is also irreducible.
In other words, we have the following short exact sequence:
\be
0\ \ \rightarrow\ S_{N,h-r} \ \ \rightarrow\ \ {\cal S}_{Nh+r}\ \
\rightarrow\ \  {\tilde S}_{N+1, r}\ \ \rightarrow\ \ 0\ .
\lb{shexS}
\ee
Here the submodule is defined as
\be
S_{N,h-r} = Span \, \{\,f_\mu^{(Nh+r)}\,,\ \mu = nh+r\, , \dots , (n+1)h-1\, \}_{n=0}^{N-1}
\lb{Sh-r}
\ee
(Nichols' series corresponding to $S_{2J+1, 1}$),
while the subquotient is
\be
{\tilde S}_{N+1, r} \simeq Span \, \{\,f_\nu^{(Nh+r)}\,,\ \nu = mh\, , \dots , mh+r-1\, \}_{m=0}^{N}\ .
\lb{Sr}
\ee
These results have been derived in cf. \cite{FHT7} by inspection of the explicit expressions (\ref{B1B2}) for the elements of the braid matrices, taking into account Lusztig's formula \cite{L}
\be
\left[ {Mh+\a\atop Nh+\b}\right] = (-1)^{(M-1)Nh + \a N - \b M}\, \left[ {\a\atop \b}\right]
\left( {M\atop N}\right)
\lb{q-bin1}
\ee
valid for $q = e^{\pm\frac{i\pi}{h}}$ and $M\in {\Z}\,,\ N \in {\Z}_+\,,\ 0\le \a,\b\le h-1$,
in which $\left( {M\atop N}\right) \in {\Z}$ is an ordinary binomial coefficient.
As we shall show in the following section, quite a similar, but in a sense dual, structure
appears in the Fock space of the WZNW {\em zero modes} which can be naturally considered as
a module over certain ``restricted`` version of $U_q s\ell(2)$ for the same values of $q$.

\section{The Fock space of WZNW zero modes as an $\mathbf{U_q}$ module}



The braiding properties of the regular basis of KZ solutions look quite natural
in the framework of the canonically quantized WZNW model, see\cite{FHIOPT, HST}
where we have considered the case $G = SU(n)$. The chiral WZNW field operator
$g(x) = \{ g^A_\a (x) \}\,$ can be then written as a sum of tensor products
\be
g^A_\a (x) = \sum_{i=1}^n u^A_i (x) \otimes a^i_\a\,,\qquad A, i ,\a = 1 , \dots , n
\lb{gua}
\ee
of generalized elementary chiral vertex operators (CVO) and ''zero modes``, respectively.
The ''full`` two-dimensional model is assumed to be defined on the conformal space-time manifold
${\mathbb S}^1 \times {\R}^1$, so the observable 2D field is periodic in the space coordinate, while
the chiral fields in (\ref{gua}) are only quasi-periodic in the corresponding light cone variable $x$. By construction, the field $g$ has a general monodromy, $\, g(x+2\pi)=g(x) M$, while the monodromy $M_p$ of $u$ is ''diagonal``; in the classical theory, $M$ belongs to the (compact) group $G$, and $M_p$ is restricted to a maximal torus.

It looks now plausible to think of a field-theoretic representation of the operators (\ref{gua}) in a space of the type (\ref{HQFQ}), with $p$ labeling in the same time the representations of the affine algebra $\widehat{\mathfrak g}$ (where ${\mathfrak g}$ is the Lie algebra of $G$, in our case, $su(n)$) at the given level, and those of the corresponding QUEA $U_q({\mathfrak g})$. The action of $u_i$ and $a^i$ on the corresponding spaces can be described as adding a box to the $i$-th row of the Young diagram (the result being zero, if this does not produce another $su(n)$ Young diagram). For the zero modes, this formalism amounts to considering a Fock-type representation of the {\it quantum matrix algebra} ${\cal A}_q$ generated by $a^i_\a$ and by a commuttative set of operators $q^{{\hat p}_{j j+1}}\,,\ j=1,2,\dots n-1$ such that $q^{{\hat p}_{j j+1}} a^i_\a = a^i_\a\, q^{{\hat p}_{j j+1}+\d^i_j -\d^i_{j+1}}$,
and assuming that ${\hat p}_{j j+1}$ are diagonalized on ${\cal V}_p$, with eigenvalues $p_{j j+1}$ equal to the corresponding shifted highest weights ($\l_j +1$, where $\l_j$ are the Dynkin labels).

This idea does not work straightforwardly for the case of interest, when the level $k$ is a non-negative integer and, accordingly, $q=e^{\pm i\frac{\pi}{h}}\,,\ h=k+n$ is a root of unity. As one might expect, the troubles come when approaching the integrability bound (of the $\widehat{su}(n)_k$ representations); for example, the exchange of two generalized CVO $u$ involves a {\it dynamical $R$-matrix} $R(\hat p)$ which may be singular on ${\cal H}_p$ if $p$ does not obey the condition $p_{12}+ \dots + p_{n-1 n} \le h-1$.

Remarkably however, the exchange of two $g$ is always well defined, being expressed in terms of a {\it numerical} (Drinfeld-Jimbo) $R$-matrix; the zero modes
$a$ accompanying the CVO ''regularize`` the chiral field operator (\ref{gua}).
For $n=2$, where the label $p=p_{12}$ takes all positive integer values, constructing primary fields out of $g(x)$, taken as elementary ones, could explain the existence of the regular KZ solutions considered in the previous section.

The zero modes $a^i_\a$ obey the quadratic exchange relations
\be
R_{12}(\hat p) a_2 a_1 = a_1 a_2 R_{12}\
\lb{inter}
\ee
(from another point of view, the tensor square of the matrix $a$ intertwines
the exchange matrices of $u(x)$ and $g(x)$).
Eq. (\ref{inter}), together with the exchange relations between $a$ and $q^{\hat p}$
and a determinant condition ($\det a = [p]$ in the case $n=2$, to which we shall restrain in what follows),
define the matrix algebra ${\cal A}_q$. In its Fock representation, $a^2_\a$ annihilate the vacuum vector $|1,0{\cal i}$, so that the Fock space ${\cal F}_q$ is spanned by the set of vectors
\be
\lb{basis2}
|p,m{\cal i} := (a^1_1 )^m (a^1_2 )^{p-1-m} |1, 0{\cal i}\qquad\quad
(\, (q^{\hat p} - q^p) |p,m{\cal i} = 0\, )\ ,
\ee
where $p=1,2,\dots $ and $m=0, \dots ,p-1$. The commutation relations for $a^i_\a$ imply
\ba
&&a^1_1 | p , m {\cal i} =  | p+1 , m+1 {\cal i} \ ,\qquad
a^1_2 | p , m {\cal i} =  q^m | p+1 , m {\cal i}\,,\nn\\
&&a^2_1 | p , m {\cal i} =  - q^{\frac{1}{2}} [p-m-1] | p-1 , m {\cal i}\ ,\nn\\
&&a^2_2 | p , m {\cal i} =  q^{m-p+\frac{1}{2}} [m] | p-1 , m-1 {\cal i} \ .
\lb{apmn2}
\ea
The exchange relations involving the Gauss components $M_\pm$ of the monodromy $M$ can be interpreted,
following the prescriptions of \cite{FRT}, as defining relations for the QUEA $U_q=U_qs\ell(2)$, so that
the entries of $M$ can be expressed in terms of its generators. Further, the exchange relations
between $M_\pm$ and the zero modes endow $a$ with the structure of a $U_q$-tensor operator, which allows to write down the relations defining the $U_q$ representation in ${\cal F}_q$ (under the assumption that the vacuum is $U_q$ invariant; $\ \varepsilon (X)$ below is the counit defined in (\ref{HopfbU})):
\ba
&&(X - \varepsilon (X)) |1,0{\cal i} = 0\quad\forall X\in U_q\ ,\qquad
q^H |p,m{\cal i} = q^{2m-p+1} |p,m{\cal i}\ ,\nn\\
&&E |p,m{\cal i} = [p-m-1]\, |p,m+1{\cal i}\ ,\qquad F |p,m{\cal i} = [m]\,|p,m-1{\cal i}\ .\qquad\qquad
\lb{Uq2}
\ea
As it follows from its definition, the quantum matrix $a$ intertwines the monodromy $M$ and the diagonal one, $M_p a = a M$. One can explicitly check that
\be
q\,\left(\begin{matrix}q^{-{\hat p}}&0\\ 0&q^{\hat p}\end{matrix}\right)\,
\left(\begin{matrix}a^1_1&a^1_2\\ a^2_1&a^2_2\end{matrix}\right) =
\left(\begin{matrix}a^1_1&a^1_2\\ a^2_1&a^2_2\end{matrix}\right)\,
\left(\begin{matrix}\l^2 FE+q^{-H-1}&-\l F q^{H-1}\\ - \l E&q^{H-1}\end{matrix}\right)
\nn\ee
with $\l=q-q^{-1}$ holds indeed in the Fock space, by using (\ref{apmn2}) and (\ref{Uq2}).

For $q$ generic, the subspaces ${\cal V}_p$ of ${\cal F}_q$ of vectors with fixed $p$ form $p$-dimensional irreducible representations of $U_q$. For $q=^{\pm i\frac{\pi}{h}}$, however, they turn into  indecomposable, in general, modules of the restricted QUEA $\bU$ (\ref{bU}). The latter has $2h$ equivalence classes of $r$-dimensional irreducible representations, $V^\pm_r$ for $1\le r\le h$ \cite{FGST}, and ${\cal V}_p$ are partially characterized by the following formula (in which $r=0$ is also allowed) which presents them as a sum of vector spaces,
\be
{\cal V}_{Nh+r} = (N+1) V_r^{\a(N)} + N V_{h-r}^{-\a(N)}\,,\quad \a(N)=(-1)^N\,,\quad {\cal V}_0=V^\pm_0 = \{0\}
\lb{VVV}
\ee
(the structure extends to an additive Grothendieck group). More precisely, the $N+1$ representations of type $V_r^{\a(N)}$ are all submodules of ${\cal V}_{Nh+r}$, and the $N$ representations of opposite ''parity`` appear as subquotients in such a way that, in the natural ordering of the label $m$, each of them is placed between two representations of the first type. Introducing Lusztig's ''divided powers`` $E^{(s)}=\frac{E^s}{[s]!}\,,\, F^{(s)}=\frac{F^s}{[s]!}\,,\,\ s=1,2,\dots$, one easily gets from (\ref{Uq2})
\be
E^{(s)} |p,m{\cal i} = \left[{p-m-1}\atop{s}\right]|p,m+s{\cal i}\ ,\quad
F^{(s)} |p,m{\cal i} = \left[{m}\atop{s}\right]|p,m-s{\cal i}\ ,
\lb{Uqprop2}
\ee
defining thus an extension $\tU$ of $\bU$, generated by $E^{(h)}$ and $F^{(h)}$. As the latter move $m$ by $h$, they connect all the components of the same parity in (\ref{VVV}). In effect, the structure of ${\cal V}_p$ {\it as $\tU$ modules} becomes similar (not equivalent but, in a sense, dual) to that encountered in the braid group representations in the previous section. Again, ${\cal V}_r$ for $1\le r\le h$, as well as ${\cal V}_{Nh}$, are irreducible, but now each ${\cal V}_{Nh+r}$ contains an $(N+1)r$-dimensional invariant irreducible submodule $V_{N+1, r}$ such that the corresponding $N(h-r)$-dimensional quotient ${\tilde V}_{N , h-r}$ is also irreducible. This is expressed by the short exact sequence
\be
0\ \ \rightarrow\ V_{N+1, r} \ \ \rightarrow\ \ {\cal V}_{Nh+r}\ \
\rightarrow\ \  {\tilde V}_{N,h-r} \ \ \rightarrow\ \ 0\ ,
\lb{shexV}
\ee
in which the subspaces forming submodules and subquotients exchange their places with respect to (\ref{shexS}).

\section{Conclusions}


It is clear that the observed duality of braid group and quantum
group representations is not a coincidence but rather an expected
feature. However, a true understanding, in the spirit of the
Kazhdan-Lusztig duality, would require additional work
(in particular, one has to identify the relevant current algebra representations
behind the regular basis of KZ solutions).
It would be interesting to study in this approach the '' transmutation`` of symmetry (from
kernels to cohomologies of screenings, in the free field setting of \cite{FGST, S})
when going back, from the logarithmic extension of a known RCFT model, to the RCFT itself.


\section*{Acknowledgements}


The authors thank I. Todorov for critical reading of the manuscript.
L.H. thanks the organizers of the International Workshop "Lie Theory and its Application in Physics VII"
(Varna, Bulgaria, June 2007), as well as SISSA,  the Central European Initiative (CEI)
and the Trieste Section of INFN  for their hospitality and support.
The work of L.H. is supported in part by the
Bulgarian National Foundation for Scientific Research (contract Ph-1406) and by the European RTN
Forces-Universe (contract MRTN-CT-2004-005104).
P.F. acknowledges the support of the Italian Ministry of University and Research (MIUR).


\end{document}